\documentclass[12pt]{article}
\usepackage[left=1.1in, right=1.1in, top=27mm, bottom=35mm]{geometry}
\usepackage{hyperref}
\usepackage[utf8]{inputenc}
\usepackage{physics}
\usepackage{amsmath, amsfonts, amsthm, amssymb}
\usepackage{mathtools}
\usepackage{titling,lipsum}
\usepackage{ragged2e}

\newcommand{\defeq}{\vcentcolon=}
\newcommand{\eqdef}{=\vcentcolon}

\newcommand{\R}{\mathbb{R}}

\numberwithin{equation}{section}

\DeclareFontFamily{T1}{calligra}{}
\DeclareFontShape{T1}{calligra}{m}{n}{<->s*[1.44]callig15}{}
\DeclareMathAlphabet\mathcalligra   {T1}{calligra} {m} {n}
\DeclareMathAlphabet\mathzapf       {T1}{pzc} {mb} {it}
\DeclareMathAlphabet\mathchorus     {T1}{qzc} {m} {n}
\DeclareMathAlphabet\mathrsfso      {U}{rsfso}{m}{n}

\begin{document}

\begin{titlingpage}

\begin{flushright}
QMUL-PH-23-17\\
\end{flushright}

\vspace{1cm}

\centering
{\scshape\LARGE  \textbf{Exotic Spheres' Metrics and Solutions \\
via Kaluza-Klein Techniques }\par}

\vspace{1cm}
{}\par

\vspace{1cm}

{T. Schettini Gherardini\par \vspace{1cm} Centre for Theoretical Physics, School of Physical and Chemical Sciences, \\
Queen Mary University of London, 327 Mile End Road, London E1 4NS, UK \\ \vspace{1cm}
E-mail: \href{mailto:t.schettinigherardini@qmul.ac.uk}{t.schettinigherardini@qmul.ac.uk}
}

\vspace{2cm}

\justifying

{ABSTRACT:  By applying an inverse Kaluza-Klein procedure, we provide explicit coordinate expressions for Riemannian metrics on two homeomorphic but not diffeomorphic spheres in seven dimensions. We identify Milnor's bundles, among which ten out of the fourteen exotic seven-spheres appear (ignoring orientation), with non-principal bundles having homogeneous fibres. Then, we use the techniques in \cite{10.1063/1.525753} to obtain a general ansatz for the coordinate expression of a metric on the total space of any Milnor's bundle. The ansatz is given in terms of a metric on $S^4$, a metric on $S^3$ (which can smoothly vary throughout $S^4$), and a connection on the principal $SO(4)$-bundle over $S^4$. As a concrete example, we present explicit formulae for such metrics for the ordinary sphere and the Gromoll-Meyer exotic sphere. Then, we perform a non-abelian Kaluza-Klein reduction to gravity in seven dimensions, according to (a slightly simplified version of) the metric ansatz above. We obtain the standard four-dimensional Einstein-Yang-Mills system, for which we find solutions associated with the geometries of the ordinary sphere and of the exotic one. The two differ by the winding numbers of the instantons involved.}

\end{titlingpage}

\tableofcontents

\newpage

\section{Introduction}
The construction of exotic spheres by Milnor represents one of the major results in modern differential
geometry (see the seminal work \cite{10.2307/1969983}). Such manifolds constitute a family of seven-dimensional spaces which are homeomorphic but not 
diffeomorphic to $S^7$, and they were first discovered as total spaces of non-trivial 3-sphere bundles over $S^4$. \\
Since the advent of general relativity, essentially all areas of theoretical physics have been permeated by differential
geometry to some extent. Motivated by the crucial role played by the metric in any physical theory, the
geometry of a manifold is usually what physicists tend to focus on. The topological characterisation, 
despite its spreading in the physics literature happened slightly later and with less rapidity, is 
also very present in current research. What seems to be missing is the intermediate layer between the two:
differentiable structures. \\
The appearance of exotic spheres in the physics literature is very rare. Shortly after their discovery,
they have been discussed by \cite{FREUND1985263} and \cite{YAMAGISHI198447} in the context of Yang-Mills instantons, although the discussion is very brief. Since then, they were considered in the
context of gravitational instantons in \cite{Witten:1985xe} and \cite{10.1063/1.529078}. The same is true for exotic manifolds (manifolds that are pairwise homeomorphic but not diffeomorphic) in general. The first consistent efforts in exploring the role of differentiable structures in physics came from Brans, who focused on how they
might be a source for gravity (\cite{Brans:1992mj}). His steps were followed by Asselmeyer-Maluga and Król, who also conducted similar investigations (see \cite{Asselmeyer-Maluga:2017tbn}, for instance). Finally, in \cite{Schleich_1999}, the authors considered a specific family 
of exotic manifolds, within the context of
gravitational path integrals. A very peculiar fact is the (almost total) absence of exotic spheres
from the string theory literature, where numerous families of seven dimensional spaces and the possible
geometries on them have been
studied - starting from the classic dimensional reductions in the 70's and 80's (a detailed list of Freund-Rubin compactifications can be found in \cite{CASTELLANI1984429}), up to the more recent
AdS-CFT investigations (\cite{Aharony_2000}, \cite{acharya1999branes}, \cite{Fr__1999}, to mention a few). It is interesting to note that different geometries on the seven-sphere, with the standard differentiable structure, have been investigated in these two contexts, as can be seen in \cite{Awada:1982pk}, \cite{POPE1985352} and \cite{Klebanov_2009}, respectively. However, the same is not true for different differentiable structures on the topological seven-sphere. This absence was noted for instance in \cite{Coquereaux:1983kj} (comment 2 therein), and, according to \cite{book}, it is due to the lack of explicit coordinates
for exotic spheres, which prevents coordinate expressions for geometrical objects such as the metric.
This explaination seems plausible by looking at the recent mathematical literature on exotic spheres.
Several studies regarding metrics on such spaces exist, but they often consist of existence results (see \cite{boyer2004einstein} and \cite{boyer2003einstein}), and even the few constructive ones are very formal, such as \cite{10.2307/1971078}, for instance. \\
In this work, we aim at bridging this gap by providing a general expression for a natural metric on
exotic spheres, with the specific case of the Gromoll-Meyer sphere worked out in full detail.\footnote{To be precise, our construction applies to those exotic spheres which appear among the family of bundles considered by Milnor, i.e. ten out of fourteen (ignoring orientation). The Gromoll-Meyer sphere is among those. For a more complete discussion, the reader is referred to \cite{nuimeprn10073}.} We obtain it by considering the original construction by Milnor, and constructing a bundle metric via the physics-inspired techniques developed in \cite{10.1063/1.525753}. Metrics that are built using this technique are known in the mathematical literature as \textit{Kaluza-Klein metrics}, or \textit{connection metrics} (constructions of this type for exotic spheres appear in \cite{10.2307/1999745} and \cite{Durán2001}, for instance). The one that we obtain is 
natural in the sense that it generalises the metric on $S^7$ viewed as the Hopf bundle, presented in \cite{DUFF19861}, in a remarkably straightforward fashion. Finally, we perform a standard non-Abelian Kaluza-Klein reduction of Einstein theory in seven-dimensions by substituting the ansatz for the connection metric in the action. We find both solutions with the geometry of the standard seven-sphere and solutions with the geometry of the Gromoll-Meyer exotic sphere. These two results appear in section \ref{sec:Milnors_bundles} and section \ref{sec:Exotic_physics}, respectively. We note that similar results, without derivations and explicit expressions, appear in \cite{FREUND1985263}, where, as we mentioned, there is one of the earliest discussions of exotic spheres that we could find in the physics literature. Sections from 2 to 5 summarise Milnor's construction of exotic spheres, the formalism of \cite{10.1063/1.525753} for bundles with homogeneous fibres and its consistency for the special case of the standard seven-sphere's principal bundle. The reader who is interested only in the results might skip those sections.

\section{Milnor's Exotic Spheres}
In this section, we outline Milnor's construction of exotic spheres. Some pedagogical reviews on the topic are: \cite{McEnroe2016MILNORSCO}, \cite{Bognat2018MILNORSES}, \cite{Exot_world}. \\
Exotic spheres are defined as the total space of an $S^3$ (non-principal) bundle over $S^4$ with transition functions given living in $SO(4)$. By the bundle reconstruction theorem (see \cite{Nakahara:2003nw}, for instance), the minimal set of information necessary to uniquely specify the total space of a bundle consists of: $M$ (the base), $\{U_i \}$ (an atlas for the base), $F$ (the fibre), $t_{ij}$ (the transition functions, which also define the structure group). To reproduce Milnor's construction of exotic spheres we employ the isomorphism between $\mathbb{R}^4$ and quaternions, which clearly induces the one between $S^3$ and unit quaternions. With this in mind, we let
\begin{align}
    M=S^4  \, , \quad \{U_i \} = \{ (U_N, \psi_N), (U_S, \psi_S) \}, \quad F=S^3 \simeq  \mathbb{H}_* = \{z \in \mathbb{H} \, | \, \norm{z}^2=1 \},
\end{align}
where $U_N = S^4 \, \backslash \, \textrm{North Pole} $  and $\psi_N$ is the usual stereographic projection from $U_N$ to $\mathbb{R}^4 \simeq \mathbb{H}$, and analogously for $U_S$, $\psi_S$. The last piece of information consists of the transition functions, which in this case is just a single one, since there are two patches on the base. We define a family of transition functions, labelled by two integers, which will in general define inequivalent bundles:
\begin{align}
f_{h,l}: \psi_N(U_N \cap U_S) \times S^3 &\xrightarrow{} \psi_S(U_N \cap U_S) \times S^3 \nonumber \\
    (z,y) &\mapsto (\frac{1}{z}, \frac{z^h y z^l}{||z||^{h+l}}),
\end{align}
where $z \in \mathbb{H}$, $y \in \mathbb{H}_*$ and quaternion multiplication is understood as juxtaposition.
The case $h=1$, $l=0$ reduces to the usual Hopf fibration (as does $h=0$, $l=1$, just with different conventions). For generic $h$ and $l$, the bundle defined this way is not a principal one, since left and right multiplication combined produce an element of $SO(4)$. What Milnor was able to prove is that for $h+l=1$, the total space of the bundle, which we will denote by $E_{h,l}$, is homeomorphic to $S^7$. In addition to that, he showed that $E_{h,l}$ and $S^7$ cannot be diffeomorphic if $(h-l)^2 \neq 1(\bmod 7)$, implying that some total spaces are exotic manifolds.\\
To conclude this section, we make a few comments on the above transition functions. By associativity of quaternions, it immediately follows that the right action and the left action commute, and we have that
\begin{align}
    \frac{z^h y z^l}{||z||^{h+l}} = \frac{1}{||z||^{h+l}} L_{z^h} R_{z^l} y = \frac{1}{||z||^{h+l}} R_{z^l} L_{z^h} y.
\end{align}
By linearity of the quaternionic multiplication, we can assign a matrix form to the operations above. For $u=a + b \textbf{i} + c\textbf{j} + d\textbf{k}$ and $x=a' + b' \textbf{i} + c'\textbf{j} + d'\textbf{k}$,
\begin{align}
    L_u x=\left(\begin{array}{cccc}
a & -b & -c & -d \\
b & a & -d & c \\
c & d & a & -b \\
d & -c & b & a
\end{array}\right)\left(\begin{array}{c}
a^{\prime} \\
b^{\prime} \\
c^{\prime} \\
d^{\prime}
\end{array}\right) \, ,
\label{eq:Left_action_quat_SO(4)}
\\
 R_u x=\left(\begin{array}{cccc}
a & -b & -c & -d \\
b & a & d & -c \\
c & -d & a & b \\
d & c & -b & a
\end{array}\right)\left(\begin{array}{c}
a^{\prime} \\
b^{\prime} \\
c^{\prime} \\
d^{\prime}
\end{array}\right) \, .
\end{align} 
If $u$ is a unit quaternion, then both matrices belong to $SO(4)$, and one can verify that the they indeed commute in general.
Specifically, $\{ L_u | u \in \mathbb{H}_* \}$ is referred as the subgroup of \textit{left-isoclinic rotations}, which we will label as $SU(2)_L$. Similarly, $\{ R_u | u \in \mathbb{H}_* \}$ are known as the \textit{right-isoclinic rotations} and this subgroup will be indicated by $SU(2)_R$. These two subgroups are not disjoint: they share the identity and the central inversion. This is simply an illustratration of the fact that $SU(2) \times SU(2)$ double covers $SO(4)$, from a slightly different angle than the usual one.

\section{Bases for so(4) and Quotients}
In this section we introduce two choices of basis for the Lie algebra of $\mathrm{so(4)}$ that will be used throughout this work. We discuss the subalgebras that naturally appear in each basis, the corresponding subgroups and the quotient spaces associated to those subgroups. 

\subsection{Rotations plus "boosts"}
\label{sec:Rot_and_boosts}
\label{sec:Rot_and_boost}
Given the usual basis $(L_{\alpha \beta})_{\mu \nu} = \delta_{\alpha \mu} \delta_{\beta \nu} - \delta_{\alpha \nu} \delta_{\beta \mu}$, we can split the generators into 3-dimensional rotations and "boosts" by defining $R_i = \frac{1}{2} \epsilon_{ijk} L_{jk}$ and $B_i = L_{i4}$, respectively. We will refer to the set of generators corresponding to this choice of basis as $\{T^{RB}_I \}$ ($I=1,...,6$). The subalgebra spanned by $\{R_i \}$, once exponentiated, produces the $SO(3)$ subgroup of $SO(4)$ which leaves the point $(0,0,0,1)\in \mathbb{R}^4$ fixed. The $B_i$'s
do not close to form a subalgebra. This can be seen from the commutation relations:
\begin{align}
    [R_i, R_j] = - \epsilon_{ijk} R_k \, , \quad  [R_i, B_j] = - \epsilon_{ijk} B_k \, , \quad [B_i, B_j] = - \epsilon_{ijk} R_k \, .
\end{align}
If we quotient by the $SO(3)$ above, we obtain $SO(4)/SO(3) \simeq S^3$, where the isomorphism is given by the map $SO(4) \ni A \mapsto Az \in S^3$ (see \cite{zbMATH03194988}), with $z=(0,0,0,1)$.
 
\subsection{Two su(2)'s}
\label{sec:Two_su(2)s}
We can define a new basis for $\mathrm{so(4)}$ by taking linear combinations of the previous generators as: $M_i = (R_i + B_i) $ and $N_i = (R_i - B_i) $.
We denote this set of generators as $\{ T^{su(2)}_I\}$. The change of basis just described illustrates the well-known Lie algebra isomorphism between $\mathrm{so(4)}$ and $\mathrm{su(2) \oplus su(2)}$. Specifically, the subset $\{ M_i \}$ spans an $\mathrm{su(2)}$ subalgebra, referred as $\mathrm{su(2)_L}$. The subset $\{ N_i \}$ spans the other $\mathrm{su(2)}$ subalgebra, referred as $\mathrm{su(2)_R}$. They exponentiate exactly to the subgroups $SU(2)_L$ and $SU(2)_R$ introduced in the previous section, respectively. The subalgebras explicitly read:\footnote{Note that the normalisation employed here is not the conventional one, but we find it to be more natural in the context of our construction.}
\begin{align}
    [M_i, M_j] = - 2 \epsilon_{ijk} M_k \, \quad [N_i, N_j] = -  2 \epsilon_{ijk} N_k .
\label{eq:Structure_constants_su(2)s}
\end{align}
We note that taking the quotient by one of these subgroups yields a different manifold from the previous case, i.e. $SO(4)/SU(2)_{L,R} \simeq SO(3)$. The reason for this is that the two subgroups have a common $\mathbb{Z}_2$ subgroup, as mentioned in the previous section.

\section{The Formalism}
\label{Sec:The_formalism}
This section outlines the geometrical setting considered by \cite{10.1063/1.525753}, the techniques used therein and their main result for us, i.e. a general expression for the metric on a bundle with homogeneous fibre.

\subsection{Conventions and Basic Notions}
\label{sec:Conventions}
Here we summarise the conventions used and the main results about Lie groups and quotient spaces that will be needed. For a more complete discussion, we refer the reader to \cite{10.1063/1.525753}. \\
Let $G$ be a Lie group and $X=G/H$ be the coset space obtained by quotienting $G$ by the subgroup $H$.
We can define a basis $\{T_i \}$ of the Lie algebra $\mathfrak{g}$ of G, with $i=1,...,\textrm{dim}(G)$ where the first $\textrm{dim}(H)$ indices span the Lie sub-algebra $\mathfrak{h} \subseteq \mathfrak{g}$ of $H$. We will denote those indices as $\hat{i}, \hat{j},...=1,...,\textrm{dim}(H)$. For the remaining indices, spanning the complementary space $\mathfrak{b} = \mathfrak{g} - \mathfrak{h}$, will use $\alpha, \beta, ...$. 
\\
A key property which we will assume is that the coset space is reductive, i.e.
\begin{align}
\textrm{Ad}_{\mathfrak{g}}(H) \mathfrak{b} \subseteq \mathfrak{b} \quad , \quad \textrm{i.e.} \quad h \mathfrak{b} h^{-1} \subseteq \mathfrak{b} \quad \forall h \in H.
\end{align}
This implies that $C_{\hat{i} \alpha}{}^{\hat{j}} = 0$, and both conditions are always satisfied if $H$ is compact. \\
Let us now make a few considerations at the level of the group (manifold).
Clearly we have a left action of $G$ on itself, given by $L_g h = g \cdot h$, and similarly a right action given by $R_h g = g \cdot h$. Moreover, $G$ has a transitive and effective left action on $X$ that we will denote by $\bar{L}: G\times X \xrightarrow{} X$. \\
Let us recall that any Lie group admits a set of left-invariant vector fields (generating right-translations), i.e.
\begin{align}
     e_i^{R}(g)=\left.\frac{d}{d t} R_{\exp \left(t T_i\right)}(g)\right|_{t=0},
\end{align}
and a set of right-invariant vector fields (generating left-translations), i.e.
\begin{align}
     e_i^{L}(g)=\left.\frac{d}{d t} L_{\exp \left(t T_i\right)}(g)\right|_{t=0}.
\end{align}
They satisfy 
\begin{align}
    \begin{aligned}
& {\left[e_i^{\mathrm{R}}, e_j^{\mathrm{R}}\right]=C_{i j}{ }^k e_k^{\mathrm{R}},} \\
& {\left[e_i^{\mathrm{L}}, e_j^{\mathrm{L}}\right]=-C_{i j}{ }^k e_k^{\mathrm{L}}.}
\end{aligned}
\end{align}
We can push forward the vector fields onto $X$ to obtain (given some coordinates $y$ on $X$):
\begin{align}
    K_i(y)=\left.\frac{d}{d t} \bar{L}_{\exp \left(t T_i\right)}(y)\right|_{t=0},
    \label{eq:K_i_definition}
\end{align}
which then obey
\begin{align}
    [K_i,K_j]=-C_{ij}{}^k K_k.
\end{align}
The $K_{\alpha}$'s form a basis on the neighbourhood of the origin of the coset space. Thus, in a neighbourhood of $O$, we have that
\begin{align}
    [K_{\alpha},K_{\beta}]=-C_{\alpha \beta}{}^i K_i{}^{\gamma}(y) K_{\gamma},
    \label{eq:K_matrix_commutator}
\end{align}
where $K_i{}^{\gamma}$ are the components of $K_i$ in the basis given by $\{K_{\alpha}\}$. Clearly, $K_{\alpha}{}^{\gamma}= \delta_{\alpha}{}^{\gamma}$, while $K_{\hat{i}}{}^{\gamma}(y)$ will be more general functions of $y$.

\subsection{Notions on Principal and Associated Bundles}
Let us consider a principal bundle $P$ with projection $\pi$, base $M$ and fibre $G$. We will refer to the coordinates on the base as $x$. Let us denote a local trivialisation by $\psi: U \times G \xrightarrow{} \pi^{-1}(U)$. We define $\phi_x : G \xrightarrow{} \pi^{-1}(U)$ as $\phi_x(g)= \psi(x,g)$. \\
Let us now construct the bundle $E$ associated with $P$, with fiber $G/H$. We will denote the projection map on $E$ as $\eta$ and the local trivialisation as $\bar{\psi}: U \times G/H \xrightarrow{} \eta^{-1}(U)$. For details on how these are constructed, we refer the reader to \cite{10.1063/1.525753}, or to the classic reference \cite{zbMATH03194988}. In the latter, they prove that $E$ is the quotient of $P$ by the right action of $H$, and we can define this quotient map as $\tau: P \xrightarrow{} E$.
Finally, if we define $\bar{\phi}_x$ analogously to $\phi_x$, we obtain the following commuting diagram
\begin{align}
    \begin{array}{ccc}
P \supseteq \pi^{-1}(x) & \stackrel{\phi_x}{\longleftarrow} & G \\
\tau \downarrow & & \downarrow \mu \\
E \supseteq \eta^{-1}(x) & \stackrel{\bar{\phi}_x}{\longleftarrow} & G / H ,
\end{array}
\label{eq:Commuting_diagram}
\end{align}
where clearly $\mu$ is the quotient map on $G$.
This diagram summarises our construction and the relation between the principal and associated bundles. \\
We can obtain a basis for vertical vectors in $E$ by considering
\begin{align}
    \bar{e}_i(w)=(\bar{\phi}_x)_* K_i(y),
    \end{align}
where $w=\bar{\psi}(x,y)$ and we are using $y$ for the coordinates on the fibre.
As before, the subset $\{\bar{e}_{\alpha} \}$ can be chosen as the set of basis vectors, satisfying:
\begin{align}
    [\bar{e}_{\alpha},\bar{e}_{\beta}]=-C_{\alpha \beta}{}^i K_i{}^{\gamma} \bar{e}_{\gamma}.
\end{align}

\subsection{General Expression for the Metric on the Total Space}
To avoid inserting a redundant section and get straight to our results, we refer the reader to section VII of \cite{10.1063/1.525753} for details on the steps leading to the gauge-invariant bundle metric on the total space. We just summarise the main points below. \\ 
The key properties imposed on the metric $\bar{g}$ on the total space $E$ (associated to some principal bundle $P$) are the following. We require it to be independent of the choice of trivialisation, to ensure gauge invariance. We also impose that $\bar{g}$ is such that horizontal and vertical spaces are orthogonal to each other. The most general expression for a metric satisfying these requirements, in the basis outlined above, is:\footnote{We recall that $x$ are the coordinates on the base and $y$ are the coordinates on the fibre.}
\begin{align}
\bar{g}_{MN} =
    \left(\begin{array}{cc}
g_{\mu \nu}(x)+\bar{h}_{\alpha \beta}(x, y) K_i{}^\alpha(y) K_j{}^\beta(y) A_\mu^i(x) A_\nu^j(x) & A_\mu^i(x) K_i{}^\alpha(y) \bar{h}_{\alpha \beta}(x, y) \\
\bar{h}_{\alpha \beta}(x, y) A_{\nu}^i(x) K_i{}^\beta(y) & \bar{h}_{\alpha \beta}(x, y)
\end{array}\right),
\label{eq:General_metric}
\end{align}
where $g_{\mu \nu}(x)$ is some metric on the base space, $A^i_{\mu}(x)$ is the component-form of some connection on the principal bundle $P$ and $\bar{h}_{\alpha \beta}(x,y)$ is the metric on $G/H$, in non-coordinate basis, and it is allowed to smoothly vary from fibre to fibre. To be more specific,
\begin{align}
\bar{h}_{\alpha \beta}(x,y) = \bar{h}(x)(K_{\alpha},K_{\beta})|_y.
\label{eq:Fibre_metric_def}
\end{align}
Hopefully this construction will become clearer with the following concrete examples.

\section{Recovering the Ordinary $S^7$ - Trivial Case}
In this section, we inspect the limits in which the metric \ref{eq:General_metric} reduces to the known metric on $S^7$, built via an inverse Kaluza-Klein construction. This allows us to introduce all the elements needed to complete the generalisation to $E_{h,l}$, which is discussed in the next section.

\subsection{Metric on $S^7$ via Inverse Kaluza-Klein à la \cite{DUFF19861}}
We now briefly review the construction of the round metric on $S^7$ viewed as a principal bundle via an inverse Kaluza-Klein process. We will repeat the steps of \cite{DUFF19861}, but with right-invariant objects instead of left-invariant ones. The equivalence between these two choices is discussed in the appendix, section \ref{Sec:Left_inv_vs_right_inv}.\\
We can write the line element defined by the usual round metric on $S^4$ as
\begin{align}
    ds^2(S^4)= d\mu^2 + \frac{1}{4}\sin^2\mu ( \bar{\Sigma}_i \bar{\Sigma}_i ),
\label{eq:Metric_on_S4_right_inv}
\end{align}
where $\bar{\Sigma}_i$ form a set of right-invariant one-forms on $S^3$ and $0 \leq \mu \leq \pi $. Clearly, this is the metric on the base space, and it is natural to also express $\bar{\Sigma}_i$ with a set of (three) angular coordinates. This is achieved by parametrising $U\in SU(2)$ with Euler angles, computing the Maurer-Cartan form $dU U^{-1} $ and decomposing it in some basis for $\mathrm{su(2)}$. Some details on how this is done in \cite{DUFF19861} are summarised in section \ref{Sec:Left_inv_vs_right_inv}.  We use $\bar{\Sigma}_i$ and $\bar{\sigma}_i$ only when referring to the right-invariant objects in the conventions of \cite{DUFF19861}. We reserve other symbols for the right/left invariant objects defined by in our conventions, which appear in the following sections. To avoid confusion of any sort, we also use a different set of Greek letters for the Euler angles: $\{ \alpha, \beta, \gamma \}$ are employed for the objects defined in our conventions, while $ \{ \phi, \theta, \psi \}$ are the letters employed in \cite{DUFF19861}.\\
Regardless of the details of the three implicit coordinates in \ref{eq:Metric_on_S4_right_inv}, the standard gauge potential for the $k=-1$ $SU(2)$ instanton centered at the origin takes the form
\begin{align}
    A^i= \cos^2(\frac{1}{2} \mu) \bar{\Sigma}_i.
    \label{eq:Instanton_gauge}
\end{align}
By \textit{standard} we mean that the gauge field is valued in $\mathrm{su(2)}$ with generators $T_i$ conventionally chosen so that $[T_i, T_j ] = \epsilon_{ijk} T_k$. This is also discussed in the appendix (see \ref{sec:Appendix_instanton}), together with the derivation of the above equation. It will shortly become evident that
\ref{eq:Instanton_gauge} is the local expression for the connection of the principal bundle defined by the Hopf fibration. In fact, by choosing the vielbein on the fibre to be another set of right-invariant one-forms $\bar{\sigma}_i$ on $S^3$, we can construct a metric on the total space as
\begin{align}
    ds^2(S^7) = d\mu^2 + \frac{1}{4} \sin^2\mu ( \bar{\Sigma}_i \bar{\Sigma}_i ) + (\bar{\sigma}_i - A^i)^2.
    \label{eq:Metric_on_S7_right_inv}
\end{align}
By the same argument of \cite{DUFF19861} for the left-invariant case, this is exactly the usual round metric on $S^7$, as anticipated. When $A^i = 0$, this the usual round metric on $S^4$ with unit radius times a  round metric $S^3$ with radius $2$. \\
The coordinates employed here are useful for the case of a single anti-instanton that we are considering, but we find them to be unsuitable for the generalisation to $|k|>1$ instantons. Hence, we report here the same quantities in stereographic coordinates:
\begin{align}
    ds^2(S^4) = \frac{4 dx^\mu dx^\mu}{(1 + x^{\mu} x^{\mu})^2} \quad \mathrm{and} \quad A^i =  \frac{1}{x^\mu x^\mu +1 } \bar{\Sigma}_i.
\label{eq:Metric_and_instanton_in_stereographic}
\end{align}
From this expression we immediately see that the moduli of the instanton have been chosen to be trivial: the centre is at the origin and the size is set to one. \\
The right-invariant one-forms $\bar{\sigma_i}$ and $\bar{\Sigma}_i$ also admit different (but equivalent) descriptions. We can define them globally when $S^3$ is thought as embedded in $\R^4$, or locally by using Euler angles as coordinates on $S^3$. While the latter is the one used in \cite{DUFF19861}, we find the former more natural in the formalism outlined in section \ref{Sec:The_formalism}. Hence, in what follows, we will use both of them and comment on the connection between the two.

\subsection{Obtaining the same Metric with the Formalism of \cite{10.1063/1.525753}}
The simplest limit of the formula \ref{eq:General_metric} is the case of $H =e$, where the fiber becomes the group itself and we recover a principal bundle. We further specialise to $G=SU(2)$ and $M=S^4$, so that $E=P$ correspond to the family of bundles containing the usual Hopf fibration of the seven-sphere. \\
Since we take $H$ to be trivial, the hatted indices disappear, and we have that $i,j,...=\alpha, \beta,... = 1, ..., \textrm{dim}(SU(2))=3$. This implies that $K_i{}^{j}(y)=\delta_i{}^j$, and consequently the metric in \ref{eq:General_metric} takes the form:
\begin{align}
    \left(\begin{array}{cc}
g_{\mu \nu}(x)+\bar{h}_{ij}(x, y) A_\mu^i(x) A_\nu^j(x) & A_\mu^i(x)  \bar{h}_{i j}(x, y) \\
\bar{h}_{i j}(x, y) A_{\nu}^j(x)  & \bar{h}_{i j}(x, y)
\end{array}\right).
\end{align}
Note that, under these assumptions, the $K_i$'s are simply the right-invariant vector fields $e_i^{L}$, which we denote with $\bar{\sigma}_i$, to connect with the previous section.
If we set $\bar{h}_{i j}(x, y) = \delta_{ij}$, then we get:\footnote{We gloss over conventions and normalisation factors in this section. They will be discussed in detail in the next section, which is a more general set-up that includes this one as a special case.}
\begin{align}
\left(\begin{array}{cc}
g_{\mu \nu}(x)+  A_\mu^i(x) A_\nu^i(x) & A_\mu^i(x)   \\
 A_{\nu}^j(x)  & \delta_{i j}
\end{array}\right).
\label{eq:Principal_bundle_metric}
\end{align}
We only need to choose $g_{\mu \nu}$ and $A_{\mu}$ at this point. We let $g_{\mu \nu}$ be the metric round metric on $S^4$ given by \ref{eq:Metric_on_S4_right_inv}. As for $A^i_{\mu}$, we set it to be the potential corresponding to a $k=-1$ instanton, i.e. $A^i = -\cos^2(\frac{1}{2} \mu) \bar{\Sigma}_i$. With these choices, we recover the line element of \ref{eq:Metric_on_S7_right_inv}. \\
This formalism for the case of the bundle being principal has already been studied thoroughly in the literature (see for instance \cite{10.1063/1.522434} and \cite{PhysRevD.13.235}), and it serves us just as the starting point for the generalisation presented in the following section.

\section{Milnor's Bundles - Non-trivial Case}
\label{sec:Milnors_bundles}
In this section, we consider Milnor's bundles as total spaces of bundles with homogeneous fibres, defined by the transition functions $f_{h,l}$. We obtain explicit expressions for the metric on bundles of the type $(0,1)$, which is diffeomorphic to the standard $S^7$, and $(-1,2)$, which is homeomorphic to $S^7$, but carries an inequivalent differentiable structure.

\subsection{Milnor's Bundles as Associated Bundles}
\label{sec:Milnors_bundles_as_associated_bundles}
Let us consider principal $SO(4)$-bundles over $S^4$.\footnote{Note that exotic spheres constructed as associated bundles to $SO(4)$ principal bundles already appeared in \cite{Rigas1978} and \cite{BOUWKNEGT201546}, for instance.} As usual, they are classified by the "winding" of their transition functions, given by $\pi_3(SO(4))$. Using the lifting theorem, we have that $\pi_3(SO(4)) = \pi(S^3 \times S^3 ) = \mathbb{Z}\times \mathbb{Z}$. The two integers correspond to $h,l$ if we choose the transition functions to be:
\begin{align}
    t_{h,l}: \psi_N(U_N \cap U_S) \times SO(4) &\xrightarrow{} \psi_S(U_N \cap U_S) \times SO(4) \nonumber \\
    (z,g) &\mapsto (\frac{1}{z}, \frac{1}{||z||^{h+l}} L_{z^h} R_{z^l} g).
\end{align}
Note that this is essentially the same choice of transition functions for the Milnor's construction, but for the object that they acting on. In other words, the only difference is the fibre. \\
It is known that the associated fibre bundle has the same transition functions of the principal one, which means that if we consider for $E$ we will have that
\begin{align}
    t_{h,l}: \psi_N(U_N \cap U_S) \times SO(4)/SO(3) &\xrightarrow{} \psi_S(U_N \cap U_S) \times SO(4)/SO(3) \nonumber \\
    (z,g) &\mapsto (\frac{1}{z}, \frac{1}{||z||^{h+l}} L_{z^h} R_{z^l} a),
\end{align}
where $a$ is some element of the quotient $SO(4)/SO(3)$ described in section \ref{sec:Rot_and_boost}. Moreover, the action of $SO(4)$ elements preserves the isomorphism between $SO(4)/SO(3)$ and $S^3$. This shows that the transition functions above indeed describe Milnor's bundles.

\subsection{The Coset Space $SO(4)/SO(3)$}
\label{sec:SO3_basis}
We now set $G=SO(4)$ and $M=S^4$. To obtain $S^3$ as the quotient space, we need $H=SO(3)$. Hence, the natural basis for the Lie algebra is given by $\mathfrak{g}$ of $G$ is given by $\{R_i, B_i \}$, defined in section \ref{sec:Rot_and_boost}. We let $\mathfrak{h}=\textrm{Span}(\{R_{\hat{i}} \})$, with $\hat{i},\hat{j}=1,2,3$ being the indices associated to this subalgebra. It follows that $\mathfrak{b}=\textrm{Span}(\{B_{\alpha} \})$, with the corresponding indices $\alpha, \beta=1,2,3$. This choices satisfies $C_{\hat{i} \alpha}{}^{\hat{j}} = 0$, which ensures that the algebra is reductive. \\
As opposed to the previous case with $H=e$, we now have that the $K_i$'s are not simply the $e_i^L$'s, and also that $K_{i}^{\alpha}$ is non-trivial. Regarding the former, we recall that this setting is special in the sense that the homogeneous space which we take to be the fibre turns out to be a Lie group again. By using \ref{eq:K_i_definition} we can find the $K_i$'s via a quick computation.
We perform the calculation by considering $S^3$ as embedded in $\mathbb{R}^4$ as usual ($S^3 = \{(X,Y,Z,W) \,\,\, \mathrm{ s.t. } \,\,\,  X^2 + Y^2 + Z^2 + W^2 = 1 \}$), and we find that the components of the $K_{\hat{i}}$'s are
\begin{align}
    (K_1)^{\mathrm{C}} = \left(\begin{array}{c}
0 \\
Z \\
-Y \\
0
\end{array}\right)^{\mathrm{C}} \, , \quad  (K_2)^{\mathrm{C}} = \left(\begin{array}{c}
-Z \\
0 \\
X \\
0
\end{array}\right)^{\mathrm{C}} \, , \quad  (K_3)^{\mathrm{C}} = \left(\begin{array}{c}
Y \\
-X \\
0 \\
0
\end{array}\right)^{\mathrm{C}} \end{align}
while the  components of the $K_{\alpha}$'s are given by
\begin{align}
(K_4)^{\mathrm{C}} = \left(\begin{array}{c}
W \\
0 \\
0 \\
-X
\end{array}\right)^{\mathrm{C}} \, , \quad  (K_5)^{\mathrm{C}} = \left(\begin{array}{c}
0 \\
W \\
0 \\
-Y
\end{array}\right)^{\mathrm{C}} \, , \quad  (K_6)^{\mathrm{C}} = \left(\begin{array}{c}
0 \\
0 \\
W \\
-Z
\end{array}\right)^{\mathrm{C}},
\label{eq:Ks_for_S3}
\end{align}
We thus have that (see equation \ref{eq:K_matrix_commutator}):
\begin{align}
K_{\hat{i}}{}^{\gamma} =
    \left(\begin{array}{ccc}
    \vspace{0.1cm}
0 & \frac{Z}{W} & -\frac{Y}{W}\\ \vspace{0.1cm}
-\frac{Z}{W} & 0 & \frac{X}{W}  \\
\frac{Y}{W} & - \frac{X}{W} & 0 \\
\end{array}\right)_{\hat{i}}^{\,\,\, \gamma}
\label{eq:K_rot_and_boost}
\end{align}
We need to be able to compare our results with the ones appearing in the literature (\cite{DUFF19861} specifically), and we also seek a convenient description for exotic spheres. For these reasons, we note that the basis just presented (which is not right-invariant) is not the most suitable one. This leads us to the next section.

\subsection{Change of Basis}
\label{sec:Change_basis}
Let us denote with $\{\mathchorus{K}_{\, \, \hat{i}'} \}$ the left-invariant vector fields on $S^3$ ($\hat{i}' = 1,2,3$), and with $\{ \mathchorus{K}_{\, \, \alpha'} \}$ the right-invariant ones ($\alpha'=4,5,6$), and it is known that each set forms a parallelisation of $S^3$. We employ the same index notation introduced in section \ref{sec:Conventions}, and the reason for this will become apparent shortly. If again we consider the unit 3-sphere as embedded in $\R^4$ with coordinates $\{X,Y,Z,W\}$, once these vector fields are normalised, their components read:
\begin{align}
    (\mathchorus{K}_{\, \, \hat{i}'})_C =  \eta^{\hat{i}'}_{CB} X_B \, , \quad (\mathchorus{K}_{\, \, \alpha'})_C=  \bar{\eta}^{\alpha'}_{CB} X_B ,
\label{eq:Left_right_inv_vector_fields}
\end{align}
where $\eta^{\hat{i}'}_{BC}$ and $\bar{\eta}^{\alpha'}_{BC}$ the self-dual 't Hooft symbols and the anti-self-dual 't Hooft symbols, respectively.
They coincide with the $K_i$'s defined by equation \ref{eq:K_i_definition} when we choose the $\mathrm{su(2) \oplus su(2)}$ basis for $\mathrm{so(4)}$, splitting the generators as $\{ M_{\hat{i}'} \}$ and $\{ N_{\alpha'} \}$:
\begin{align}
    \mathchorus{K}_{\,\, \hat{i}'}(y)=\left.\frac{d}{d t} \bar{L}_{\exp \left(t M_{\hat{i}'}\right)}(y)\right|_{t=0} \, , \quad 
    \mathchorus{K}_{\,\, \alpha'}(y)=\left.\frac{d}{d t} \bar{L}_{\exp \left(t N_{\alpha'}\right)}(y)\right|_{t=0} .
\end{align}
Just as a quick check, we observe that they satisfy
\begin{align}
    [\mathchorus{K}_{\,\, \hat{i}'}, \mathchorus{K}_{\,\, \hat{j}'}] = 2 \epsilon_{\, \hat{i}' \hat{j}' \hat{k}'} \mathchorus{K}_{\,\, \hat{k}'} \, , \quad [\mathchorus{K}_{\,\, \alpha'}, \mathchorus{K}_{\,\, \beta'}] = 2 \epsilon_{\, \alpha' \beta' \gamma'} \mathchorus{K}_{\,\, \gamma'},
\end{align}
which is the opposite of \ref{eq:Structure_constants_su(2)s}, as we would expect from section \ref{Sec:The_formalism}.
We note that
\begin{align}
    \mathchorus{K}_{\,\, 1} = K_1 + K_4 \, \quad \mathchorus{K}_{\,\, 2} = K_2 + K_5 \, \quad \mathchorus{K}_{\,\, 3} = K_3 + K_6 \, , \\ 
    \mathchorus{K}_{\,\, 4} = K_1 - K_4 \, \quad \mathchorus{K}_{\,\, 5} = K_2 - K_5 \, \quad \mathchorus{K}_{\,\, 6} = K_3 - K_6 \, .
\end{align}
It can be summarised by
\begin{align}
    \mathchorus{K}_{\,\, i'}=M_{i'}{}^{i} K_i ,
\end{align}
with
\begin{align}
M_{i'}{}^{i} =
\left(\begin{array}{cccccc}
1 & 0 & 0 & 1 & 0 & 0 \\
0 & 1 & 0 & 0 & 1 & 0 \\
0 & 0 & 1 & 0 & 0 & 1 \\
1 & 0 & 0 & -1 & 0 & 0 \\
0 & 1 & 0 & 0 & -1 & 0 \\
0 & 0 & 1 & 0 & 0 & -1 
\end{array} \right)_{i'}^{\,\,\,\,\, i} 
\end{align}
with inverse given by
\begin{align}
    K_i = (M^{-1})_{i}{}^{i'} \mathchorus{K}_{\,\, i'},
\end{align}
so that 
\begin{align}
    M_{i'}{}^i (M^{-1})_{i}{}^{j'} = \delta_{i'}^{j'} \quad \textrm{and} \quad (M^{-1})_{i}{}^{i'} M_{i'}{}^{j} = \delta^{j}_{i}.
\end{align}
The same holds for the basis at the origin clearly, i.e. the $T^{su(2)}$'s and $T^{RB}$'s are related by $T^{su(2)}_{ i'}=M_{i'}{}^{i} T^{RB}_i$. 
Now, this means that
\begin{align}
    \bar{h}_{\alpha' \beta'} = \bar{h}(\mathchorus{K}_{\,\, \alpha'}, \mathchorus{K}_{\,\, \beta'}) = \bar{h}(M_{\alpha ' }{}^{i} K_{i}, M_{\beta ' }{}^{j} K_{j}) = M_{\alpha ' }{}^{i} M_{\beta ' }{}^{j} \bar{h}(K_{i}, K_{j}) = \nonumber \\ M_{\alpha ' }{}^{i} K_i{}^{\alpha} M_{\beta ' }{}^{j} K_j{}^{\beta}  \bar{h}_{\alpha \beta} = W_{\alpha'}{}^{\alpha} W_{\beta'}{}^{\beta}  \bar{h}_{\alpha \beta}, 
\end{align}
where we have implicitly defined $W_{\alpha'}{}^{\alpha} = M_{\alpha ' }{}^{i} K_i{}^{\alpha}$, with inverse $(W^{-1})_{\alpha}{}^{\alpha'}$\footnote{Note that, even though $K_i{}^{\alpha}$ does not have an inverse, its contraction with $ M_{\alpha ' }{}^{i}$, which turns it into a square matrix, does.}.
Analogously, we have that $A_{\mu}=A_{\mu}^{i'} T^{su(2)}_{i'} = A_{\mu}^{i} T_{i}^{RB}$ gives
\begin{align}
    A_{\mu}^{i} = A_{\mu}^{i'} M_{i'}{}^i . 
\end{align}
Finally, 
\begin{align}
     M_{i'}{}^{i} K_i =  M_{i'}{}^{i} K_i{}^{\alpha} K_{\alpha}  =\mathchorus{K}_{\,\, i'} = \mathchorus{K}_{\,\, i'}{}^{\alpha'} \mathchorus{K}_{\,\, \alpha'} = \mathchorus{K}_{\,\, i'}{}^{\alpha'} M_{\alpha ' }{}^{i} K_{i} =  \mathchorus{K}_{\,\, i'}{}^{\alpha'} M_{\alpha ' }{}^{i} K_i{}^{\alpha} K_{\alpha},
\end{align}
from which we infer
\begin{align}
    \mathchorus{K}_{\,\,i'}{}^{\alpha'} =   M_{i'}{}^{i} K_i{}^{\alpha} (W^{-1})_{\alpha}{}^{\alpha'} .
\end{align}
An identical reasoning can be applied to obtain $K_i{}^{\alpha}$ from $\mathchorus{K}_{\,\,i'}{}^{\alpha'}$. One can check that, using \ref{eq:K_rot_and_boost} together with $K_{\alpha}{}^{\gamma} = \delta_{\alpha}{}^{\gamma}$, one obtains
\begin{align}
    \mathchorus{K}_{\,\,\hat{i}'}{}^{\alpha'} =
\left(
\begin{array}{ccc} \vspace{0.1 cm}
 1-2 \left(W^2+X^2\right) & -2 (W Z+X Y) & 2 W Y-2 X Z \\ \vspace{0.1cm}
 2 (W Z-X Y) & 1-2 \left(W^2+Y^2\right) & -2 (W X+Y Z) \\
 -2 (W Y+X Z) & 2 W X-2 Y Z & 2 \left(X^2+Y^2\right)-1 
\end{array}
\right)_{\hat{i}'}^{\,\,\,\, \alpha '} ,
\label{eq:K_in_left_right_basis}
\end{align}
with $\mathchorus{K}_{\,\,\gamma '}{}^{\alpha'} = \delta_{\gamma '}{}^{\alpha'}$, as expected. This result will be needed in section \ref{sec:Exotic_physics}. For now, we point out that each row of the above matrix can be thought as a map from $S^3$ embedded in $\R^4$ to $S^2$ embedded in $\R^3$. Specifically, each row is a realisation of the Hopf map $p$ that defines the Hopf fibration $S^1 \hookrightarrow S^3 \xrightarrow{p} S^2$. \\
As we mentioned, the right/left-invariant basis is the most natural one to work with when dealing with Milnor's bundles. Since we just uncovered its connection with the one discussed in section \ref{sec:SO3_basis}, we will work in the left/right-invariant basis in the following sections. Under this assumption, we now drop the primes in the indices to ease the notation.

\subsection{Usual Sphere (again)}
\label{sec:Usual_sphere_2}
Armed with the results obtained in the previous two sections, we are almost in the position to recover the metric \ref{eq:Metric_on_S7_right_inv} in this new setting. We define everything in the basis of left/right-invariant objects, knowing that it is related to the natural basis of section \ref{sec:SO3_basis} via the maps defined in section \ref{sec:Change_basis} (again, we drop the primes under this assumption).\\
The components of the dual one-forms to \ref{eq:Left_right_inv_vector_fields} read (see also section \ref{Sec:left_and_right_vect_fields} for more details):
\begin{align}
    (\bar{\omega}_{ \hat{i}})_C =  \eta^{\hat{i}}_{CB} X_B \, , \quad (\bar{\omega}_{ \alpha})_C=  \bar{\eta}^{\alpha}_{CB} X_B .
    \label{eq:Dual_forms_omega}
\end{align}
Let us now make the connection between this set-up and the formalism of \cite{DUFF19861}, pointing out a few subtle differences. Firstly, we have that the norm of the $\bar{\omega}_{ \alpha}$'s is half the norm of the $\bar{\Sigma}_\alpha$'s ($\alpha = 1,2,3$). The $\bar{\omega}_{ \alpha}$'s are uniformly scaled by a factor of $1/2$ so that $|\bar{\omega}_{ \alpha}| = 1/2 \,  |\bar{\Sigma}_{\alpha}| $ for any $\alpha$.
With this consideration in mind, let us choose $\bar{h} = 4 \bar{\omega}_{\alpha} \bar{\omega}_{\alpha}$, so that, according to \ref{eq:Fibre_metric_def}, we obtain $\bar{h}_{\alpha \beta}= 4 \delta_{\alpha \beta}$.
For the metric on the base, we let
\begin{align}
    ds^2(S^4)= d\mu^2 + \sin^2\mu ( \bar{\Omega}_{\alpha} \bar{\Omega}_{\alpha} ),
\label{eq:Metric_on_S4_right_inv_2}
\end{align}
where $\bar{\Omega}_{\alpha}$ is another set of Maurer-Cartan forms identical to \ref{eq:Dual_forms_omega}.
Regarding the gauge field, we recall that our definition of generators for the algebra was unconventional (see equation \ref{eq:Structure_constants_su(2)s}). With this choice, we have that an anti-instanton in the $\mathrm{su(2)}$ subalgebra labelled by Greek indices is described by $A^{\alpha}= -\cos^2(\frac{\mu}{2}) \bar{\omega}_{\alpha}$. We let the gauge field living in the other $\mathrm{su(2)}$ be trivial, i.e. $A^{\hat{i}}=0$.\\
Now, by recalling that $\mathchorus{K}_{\,\, \alpha}{}^{\gamma}= \delta_{\alpha}{}^{\gamma}$ (see \ref{eq:K_matrix_commutator} and the following comments), we can plug all these quantities in equation \ref{eq:General_metric}. By doing this, we find that the metric on the $(0,1)$ Milnor's bundle that we just outlined coincides with \ref{eq:Metric_on_S7_right_inv}, as expected.

\subsection{The Double Instanton}
In order to construct the Gromoll-Meyer exotic sphere, we need a transition function $f_{h,l}$ with $h=2$, $l=-1$, or anything homotopic to this (see \cite{nuimeprn10073}). Translating this in the physics language, we need an anti-instanton in one $\mathrm{su(2)}$ factor and a double instanton, i.e. $k=2$, in the other factor. While for the first one there exists a simple a explicit expression involving all the moduli, the same is not true for the second one. A closed form in terms of all the moduli exists, but it is quite complicated. We will now take a small detour to clarify this point. The most common approach to finding the general solution to the (anti-)self-duality equations, which is also the one that first appeared in the literature, starts by taking the following ansatz for the gauge field:
\begin{align}
    A_{\mu} = \sigma_{\mu \nu } \partial_{\nu} ln \rho.
\end{align}
Then, it is shown that $\rho$ must satisfy $\frac{1}{\rho} \Box \rho = 0$. The general solution is found to be:\footnote{To be precise, this misses four moduli, see \cite{PhysRevD.15.1642}.}
\begin{align}
    \rho = 1 + \sum_{i=1}^k \frac{\lambda_i^2}{(x - a_i)^2},
\end{align}
and it has winding number $k$. However, there is a subtlety when $a_i = a_j$ and $i \neq j$. It is easy to see that, for the case of $k=2$, if $a_1 = a_2$, then we obtain the a $k=1$ instanton with size squared given by $\lambda_1^2 + \lambda_2^2$. This is noted in \cite{RevModPhys.51.461}, for instance, and this seemingly singular point in the moduli space is the reason why we were not able to obtain an $SO(4)$-invariant solution for the $k=2$ gauge potential.\footnote{Note that the most elegant formalism for this computation is the ADHM construction, that can be translated into explicit expressions for the gauge potential and field strength. This is work in progress. We thank Professor Berman and Professor Travaglini for discussions on this point.} With this subtle point in mind, we present here the expression for the gauge field describing two instantons ($k=2$) with the same size:
\begin{align}
    A^a_{\mu} = -2\frac{(x-a)^2 (x-b)^2}{(x-a)^2 (x-b)^2 + \rho^2 [(x-a)^2 + (x-b)^2]} \eta^a_{\mu \nu} \Bigg( \frac{\rho^2 (x-a)_{\nu}}{(x-a)^4} + \frac{\rho^2 (x-b)_{\nu}}{(x-b)^4}
     \Bigg).
     \label{eq:Diinstanton}
\end{align}
For convenience, the gauge field here is given in the \textit{singular gauge}, while the one in \ref{eq:Metric_and_instanton_in_stereographic} is in the \textit{regular gauge}. We also chose the sizes of the two instantons to be the same here in order to keep our expression general without making it too cumbersome.

\subsection{Exotic Spheres}
\label{sec:Exotic_sphere_metric}
Let us now move to the general case involving both $\mathrm{su(2)}$ components of the connection, keeping the same choices for $g_{\mu \nu}(x)$ and $\bar{h}_{\alpha \beta}(y)$. \\
Given a general connection $A_{\mu}^i$ on $P$, the metric in \ref{eq:General_metric} will include contributions from the $\mathchorus{K}_{\, \, \hat{i}}{}^{\gamma}(y)$.
In this case we will need the most general form of the metric on $E$, given by
\begin{align}
    \left(\begin{array}{cc}
g_{\mu \nu}(x) + 4\delta_{\alpha \beta} \mathchorus{K}_{\,\,i}{}^\alpha(y) \mathchorus{K}_{\,\,j}{}^\beta(y) A_\mu^i(x) A_\nu^j(x)   \,\,\,
&  4A_\mu^{\beta}(x) 
 + 4A_\mu^{\hat{i}}(x) \mathchorus{K}_{\,\,\hat{i}}{}^\beta(y)  \\
 4A_{\nu}^{\alpha}(x) + 4A_{\nu}^{\hat{i}}(x) \mathchorus{K}_{\,\,\hat{i}} {}^\alpha(y) & 4\delta_{\alpha \beta}
\end{array}\right),
\label{eq:Expansion_metric}
\end{align}
where, in the off-diagonal entries, we simply expanded $ A_{\nu}^i(x) \mathchorus{K}_{\,\,i}{}^{\alpha}$. For completeness, the expansion of $\delta_{\alpha \beta} \mathchorus{K}_{\,\,i}{}^\alpha(y) \mathchorus{K}_{\,\, j}{}^\beta(y) A_\mu^i(x) A_\nu^j(x) $ reads
\begin{align}
    A^{\alpha}_{\mu}(x) A^{\alpha}_{\nu}(x) + A^{\alpha}_{\mu}(x) \mathchorus{K}_{\,\,\hat{j}}{}^{\alpha} A^{\hat{j}}_{\nu}(x) + A^{\hat{i}}_{\mu}(x) \mathchorus{K}_{\,\,\hat{i}}{}^{\beta} A^{\beta}_{\nu}(x)  +  \mathchorus{K}_{\,\,\hat{i}}{}^\alpha(y) \mathchorus{K}_{\, \, \hat{j}}{}^\beta(y) A_\mu^{\hat{i}}(x) A_\nu^{\hat{j}}(x).
    \label{eq:Expansion_first}
\end{align}
In the previous case, by setting $A^{\hat{i}}=0$, we eliminated the two extra contributions in the off-diagonal terms of \ref{eq:Expansion_metric} and the last three terms in \ref{eq:Expansion_first}. While, in this case, we set the gauge field to be a combination of an anti-instanton (\ref{eq:Instanton_gauge}) and a double instanton (\ref{eq:Diinstanton}):
\begin{align}
    &A^{\alpha}_{\mu}= - \bar{\eta}^{\alpha}_{\mu \nu} \frac{x_{\nu}}{x^2 +1} \, , \label{eq:Single_instanton_explicit} \\ 
     &A^{\hat{i}}_{\mu} =  -\frac{(x-a)^2 (x-b)^2}{(x-a)^2 (x-b)^2 + \rho^2 [(x-a)^2 + (x-b)^2]} \eta^{\hat{i}}_{\mu \nu} \Bigg( \frac{\rho^2 (x-a)^{\nu}}{(x-a)^4} + \frac{\rho^2 (x-b)^{\nu}}{(x-b)^4}
     \Bigg). 
    \label{eq:Double_instanton_explicit}
\end{align}
This corresponds to Milnor's bundle with transition maps with winding numbers $(2,-1)$, which is an exotic sphere. To make the construction fully explicit, we should provide the matrix $\mathchorus{K}_{\,\, \hat{i}}{}^{\alpha}$ in some coordinates. We do that by choosing the Euler angles, as described in the appendix (see \ref{Sec:left_and_right_vect_fields}), which yields:
\begin{align}
    \mathchorus{K}_{\,\, \hat{i} }{}^{\alpha} = \left(\begin{array}{ccc}
    \vspace{0.1cm}
       \cos\alpha \cos \beta \cos \gamma - \sin \alpha \sin \gamma \,\, & \,\, -\cos \beta \cos \gamma \sin \alpha - \cos \alpha \sin \gamma \,\,  &  \,\, \cos \gamma \sin \beta \\ \vspace{0.1cm}
       \cos \gamma \sin \alpha + \cos \alpha \cos \beta \sin \gamma \,\,  & \,\, \cos \alpha \cos \gamma - \cos \beta \sin \alpha \sin \gamma \,\, & \,\, \sin \beta \sin \gamma \\
       \cos \alpha \sin \beta & \sin \alpha \sin \beta & -\cos \beta 
    \end{array} \right)_{\hat{i}}^{\,\,\,\,\, \alpha}  
\end{align}
Now, let us make a few remarks about the metric just presented. With the choices outlined above, this is the simplest bundle Riemannian metric on the Gromoll-Meyer sphere, in that it is written as a slight generalisation of the metric in \cite{DUFF19861}. However, this is far from being the most general bundle metric on such exotic sphere. We can generalise three ingredients in our construction as follows. Firstly, we can make a different choice for the metric on the base $S^4$. Secondly, we can consider a different metric on the fibre $S^3$, as long as it is left-invariant (see \cite{10.1063/1.525753}). And, finally, we can employ a wider class of gauge fields then the (anti)-self-dual ones. We restricted to instanton ansatzes for the $|k|=1$ and $|k|=2$ components of the $\mathrm{so(4)}$ connection, but this assumption can be relaxed. However, as described in the next section, the choice of instanton connections does play a role when the exotic geometry appears in a physical theory.

\section{Traces of Exoticness in Four Dimensions}
\label{sec:Exotic_physics}
As we argued in section \ref{sec:Milnors_bundles_as_associated_bundles}, by setting $M=S^4$, $G=SO(4)$ and $H=SO(3)$, we obtain exactly Milnor's bundles. The family of their total spaces contains the standard seven-sphere and many exotic ones. In this section, we present an instance of how two geometries associated to inequivalent differentiable structures might appear as solutions to a physical theory. To do so, we simply follow Kaluza-Klein's prescription, in our non-abelian setting. We obtain a four-dimensional theory by substituting the metric ansatz \ref{eq:General_metric} into the action of seven-dimensional gravity, with cosmological constant, and integrating over the fibre. To avoid complications, we consider the case where $\bar{h}_{\alpha \beta}$ has no $x$-dependence, as it is done in \cite{DUFF198490}, for instance. Solutions for the more general case, which includes the $x$-dependence, are currently being studied, together with their higher-dimensional interpretations in terms of D-branes. \\
The dimensional reduction outlined above yields:
\begin{align}
\begin{aligned}
\int_E d^4 x \, d^3 y  \, \bar{g}^{1 / 2} ( \bar{R} - 2 \Lambda) = V_{G/H} \int_M d^4 x \sqrt{g}  \Big(  R_M  - \frac{1}{4} g^{\mu \nu} g^{\rho \sigma} \lambda_{i j} F_{\mu \rho}{ }^i F_{\nu \sigma}{ }^j  + R_{\, G/H} -2 \Lambda \Big).
 \label{eq:Action_reduced}
 \end{aligned}
\end{align}
This result can be read off from the one in \cite{10.1063/1.525753}, by ignoring the terms coming from the $x$-dependence of the fibre metric, and we refer to section VIII therein for the intermediate steps in the computation of the reduced Ricci scalar. Using the same notation, we have defined $R_M$, $R_{G/H}$ and $V_{G/H}$ as the Ricci scalar of the base, the Ricci scalar of the fibre and the volume of the fibre, respectively. Regarding the definition of $\lambda_{ij}$, we take a small detour. It is given by:
\begin{align}
    \lambda_{ij} = \frac{1}{V_{G/H}} \int_{G/H} d^3 y \,[\bar{h}]^{1/2} \,\, \bar{h}_{\alpha \beta} K_{i}{}^{\alpha}{}(y) K_{j}{}^{\beta}{} (y) = \frac{1}{V_{G/H}} \int_{G/H} d^3 y \,[\bar{h}]^{1/2} \,\, K_{i}{}^{\alpha}{}(y) K_{j}{}^{\alpha}{} (y),
\end{align}
where $ V_{G/H} = \int_{G/H} d^3 y \,[\bar{h}]^{1/2}$. To avoid a proliferation of factors of two, we have chosen $\bar{h}_{\alpha \beta} = \delta_{\alpha \beta}$, differently from what was done in section \ref{sec:Usual_sphere_2}.\footnote{Note that this is consistent with scaling the generators by a factor of $1/2$ compared to the previous sections. Schematically, if $M,N \xrightarrow{} 1/2 M, 1/2 N$, then we have that $\mathchorus{K}_{\,\,i} \xrightarrow{} 1/2 \mathchorus{K}_{\,\,i}$, while $\mathchorus{K}_{\,\, \hat{i}}{}^{\alpha}$ stays the same. Hence, the dual forms to the $\mathchorus{K}_{\,\, \alpha}$'s are scaled by a factor of two, and so are the gauge fields $A^{\alpha}$, giving the standard expression for the anti-instanton.}
As we pointed out (see comments after equation \ref{eq:K_matrix_commutator}), when $i = \gamma$, we have that $K_{\gamma}{}^{\alpha}{} = \delta_{\gamma}{}^{\alpha}{} $, and hence $\lambda_{\alpha \beta} = \delta_{\alpha \beta}$.  For the other cases, we need to examine $K_{\hat{i}}{}^{\alpha}{}$. They are the coefficients for the change of basis between the right-invariant and the left-invariant vector fields on $S^3$. To make the symmetries manifest, and make simplifications easier, we again think of $S^3$ as embedded in $\mathbb{R}^4 = \{ (X,Y,Z,W) | X,Y,Z,W \in \mathbb{R} \}$. Then, we have that (see \ref{eq:K_in_left_right_basis}):
\begin{align}
    \begin{aligned}
        K_1 = (-W^2 - X^2 +Y^2 + Z^2 , \, -2XY -2WZ , \, 2WY -2XZ) , \\
         K_2 = (-2XY + 2WZ, \, -W^2 + X^2 - Y^2 + Z^2, \, -2WX -2YZ) , \\
         K_3 = (-2WY -2XZ, \, 2WX - 2YZ, \, -W^2 + X^2 + Y^2 - Z^2 ).
    \end{aligned}
\end{align}
As we mentioned, they are all different realisations of the Hopf map from $S^3$ to $S^2$, which is curious. We see immediately that the integral of any single component of $K_{\hat{i}}{}^{\alpha}{}$ over the three-sphere vanishes, due to symmetry. This implies $\lambda_{\alpha \hat{i}} = 0$. For the same reason, $\lambda_{\hat{i} \hat{j}} = 0$ for $\hat{i} \neq \hat{j}$. The only case where we get a non-zero integral is when $\hat{i} = \hat{j}$, where we have that $K_{\hat{i}}{}^{\alpha}{} K_{\hat{i}}{}^{\alpha}{} = 1$ (no sum over $\hat{i}$), so that $\lambda_{\hat{i} \hat{i}}=1$. \\
Given that $\lambda_{ij} = \delta_{ij}$, the dimensional reduction yields Einstein-Yang-Mills action (with cosmological constant given by the Ricci scalar of the fibre), which agrees with the analogous cases examined in \cite{DUFF198490} and \cite{10.1063/1.522434}.
Hence, the dynamics of the theory is described by the standard equations of this system. Varying with respect to the metric yields: 
\begin{align}
    R_{\mu \nu} - \frac{1}{2} g_{\mu \nu} R - \frac{1}{2} g_{\mu \nu} R_{S^3} + g_{\mu \nu} \Lambda =  \frac{1}{2} \lambda_{ij} \Big[ g^{\rho \sigma} F_{\mu \rho}^i F_{\nu \sigma}^j - \frac{1}{4} g_{\mu \nu} F_{\rho \sigma}^i F^{\rho\sigma \, j} \Big] ,
    \label{eq:EoM_metric}
\end{align} 
while we do not perform the explicit variation with respect to the gauge fields because we take the Bogmonly' bound shortcut.
Firstly, we let $g_{\mu \nu}$ be the metric on $S^4$ with radius $R$, which in stereographic coordinates reads:
\begin{align}
    g_{\mu \nu} = \frac{4 R^4}{(R^2 + x^2) ^2} \delta_{\mu \nu},
\label{eq:Metric_on_S4_stereo}
\end{align}
with the determinant being $g = \big(\frac{4 R^4}{(R^2 + x^2) ^2} \big)^4$. Ricci curvature tensor, scalar curvature and Einstein tensor read:
\begin{align}
    R_{\mu \nu} = \frac{12 R^2}{(R^2 + x^2) ^2} \delta_{\mu \nu} \, ,  \quad R_{S^4} = \frac{12}{R^2} \, \, \, \implies \, \, \, G_{\mu \nu} = - \frac{12 R^2}{(R^2 + x^2) ^2} \delta_{\mu \nu} = - \frac{3}{R^2} g_{\mu \nu} \, ,
    \label{eq:Ricci_and_Einstein}
\end{align}
respectively. Then, as anticipated, we consider the self-duality equation:
\begin{align}
    F_{\mu \nu} = \frac{1}{2 \sqrt{g}} \epsilon^{\rho \sigma \tau \omega }  g_{\mu \rho} g_{\nu \sigma} F_{\tau \omega},
    \label{eq:Self_duality_curved}
\end{align}
where $\epsilon^{\rho \sigma \tau \omega } $ is the Levi-Civita symbol. The overall scaling factor in \ref{eq:Metric_on_S4_stereo} of the two contracted metrics cancels with the square root of the inverse determinant, so that \ref{eq:Self_duality_curved} reduces to the standard self-duality in $\mathbb{R}^4$, whose solutions are the well known instantons.
Hence, we can set the gauge field to be an anti-instanton:
\begin{align}
    \left( \begin{array}{c}
    A^{\alpha}_\mu \\
    A^{\hat{i}}_\mu
    \end{array} \right)
    = \left(\begin{array}{c}
    2 \bar{\eta}^{\alpha}_{\mu \nu} \frac{ x^\nu}{x^2 + R^2} \\
    0
    \end{array} \right).
\end{align}
This choice recovers the seven-dimensional ordinary sphere. Correspondingly, the choice associated to the Gromoll-Meyer sphere reads:
 \begin{align}
    \left( \begin{array}{c}
    A^{\alpha}_\mu \\
    A^{\hat{i}}_\mu
    \end{array} \right)
    =\left(\begin{array}{c}
    \mathrm{equation} \,\, \ref{eq:Single_instanton_explicit} \\
\mathrm{equation} \,\, \ref{eq:Double_instanton_explicit}
    \end{array} \right) .
\end{align}
By the usual argument, since these gauge fields satisfy the self-duality equation \ref{eq:Self_duality_curved}, then they automatically satisfy their equations of motion (see \cite{Oh:2011nv}, for example). Moreover, they give a vanishing energy-momentum tensor.
Hence, for both choices, the right-hand side of equation \ref{eq:EoM_metric} vanishes. Then, by using \ref{eq:Ricci_and_Einstein}, we find that $\Lambda = \frac{3}{R^2} + \frac{R_{S^3}}{2} $, which completes our solution.  \\
As mentioned above, we leave the discussion of the higher-dimensional field equations for a forthcoming article. However, let us make just a quick comment about the consistency of this dimensional reduction. The $y$-dependence in the 4-d equations of motion, which is the source of inconsistency pointed out in \cite{DUFF198490}, does not affect the solution. The reason for this is our choice of a four-dimensional Einstein space and an instanton gauge field, which ensures that both sides of equation \ref{eq:EoM_metric} vanish (this should be compared with equation (3) in \cite{DUFF198490}). Finally, we also note that the above article contains an argument for the existence of a consistent ansatz when the fibre is itself a non-abelian group manifold, which holds in our case due to the diffeomorphism $S^3 \simeq SU(2)$.

\section{Conclusions and Outlook}
This work is divided into two parts, excluding the background material. \\
In the first one, we discussed the construction of a metric on Milnor's bundles following an approach inspired by Kaluza-Klein theories. We reviewed the formalism of \cite{10.1063/1.525753} for bundles with homogeneous fibre and introduced the explicit expression of the most general metric on the total space compatible with the bundle structure. In the trivial limit of the fibre being diffeomorphic to the structure group, the metric is the same obtained via the standard inverse Kaluza-Klein construction. We focused on the most relevant example of this limit: $S^7$ viewed as the principal Hopf bundle, which also appeared in \cite{DUFF19861}. Then, after explicitly constructing exotic spheres as total spaces of bundles with homogeneous fibres, we applied the formalism to two of these manifolds. The first one is the seven-sphere with the ordinary differentiable structure, with the associated metric being the usual round one. The second one, which is more interesting and represents the new result of this work, is Milnor's bundle characterised by the pair of winding numbers $(2,-1)$, which carries an inequivalent differentiable structure. In this case, we find that the associated metric is a straightforward modification of the round metric on $S^7$. Simplicity is the key feature of this metric, which does not have any special properties (it is not Einstein, for instance). By simplicity, we mean that it arises naturally in the context of Kaluza-Klein formalism and its expression in coordinates is remarkably similar to the round metric on the ordinary $S^7$. However, looking for a metric with special properties in the space of all metrics compatible with the bundle structure is a very interesting direction, subject of an ongoing investigation. \\
In the second part, we studied the dimensional reduction of gravity (with cosmological constant) for manifolds belonging to the family of Milnor's bundles. Via a non-abelian Kaluza-Klein mechanism, with integration over the $S^3$ fibre, we obtained four-dimensional Einstein-Yang-Mills theory. We found explicit solutions for both manifolds of the exotic pair considered, which differ by the winding numbers of the instantons involved, showing how inequivalent differentiable structures in seven-dimensions descend to different solutions in the four-dimensional theory. To explore the full range of possible solutions - not only to the reduced theory, but also to the higher-dimensional field equations - it is necessary to use explicit expressions for $k=2$ connections with all the moduli. As we mentioned, this is part of an ongoing investigation, involving the ADHM construction of instantons and the interpretation of the solutions in the context of supergravity.  \\
As we mentioned in the introduction, the results presented are only partially novel, in that similar ideas appeared briefly in \cite{FREUND1985263}. However, neither the approach described here, nor the derivation of the explicit expressions are something that we could find in the existing literature. \\
This brings us to the possible extensions of this work. \\
The search for explicit Einstein metrics on these spaces is surely a fascinating one.
We also note that we limited ourselves to just one of the 27 exotic differentiable structures on $S^7$, to avoid cumbersome expressions. We believe that studying the results of this construction for the remaining exotic spheres would give valuable insights on such manifolds, as well as the Berger space, which was recently shown in \cite{10.2307/40067878} to be diffeomorphic to the total space of an $S^3$-bundle over $S^4$. Regarding more concrete applications in physics, we mainly point out two directions. As we mentioned, the first one consists of extending this work to finding exotic sphere solutions in 7D supergravity. A second interesting option would be to try and construct Freund-Rubin-like solutions to 11D supergravity, with an exotic sphere as the internal manifold  (note that such solutions with non-Einstein metrics on the internal space exist, see \cite{POPE1985352}, for instance). For both directions, studying the specific properties of $k=2$ instantons is a crucial step.
\newpage

\section*{Acknowledgements}
I am particularly grateful to my supervisor, Professor David Berman. For introducing me to the existence of exotic spheres, for the support and the stimulating discussions. I would also like to thank Guglielmo Tarallo, for his help with bundle theory. I am grateful Professor Johnson, Professor Schleich and Professor Derdzinski (in chronological order) for their support during the early stages of this project. Finally, I would like to thank Dr Mattia Cesaro for his comments on the manuscript. This work was funded by the Science and Technology Facilities Council (STFC) Consolidated Grants ST/T000686/1 "Amplitudes, Strings \& Duality" and ST/X00063X/1 "Amplitudes, Strings \& Duality". No new data were generated or analysed during this study.

\appendix

\section{More Details on the Geometrical Objects}
\vspace{0.3cm}
\subsection{Left- and Right-Invariant Vector Fields on $S^3$}
\label{Sec:left_and_right_vect_fields}
In this section, we review the left/right-invariant vector fields and the Maurer-Cartan forms on $S^3 \simeq SU(2)$, both with their local and global descriptions. Analogous discussions can be found in \cite{10.1063/1.2358391} and \cite{https://doi.org/10.15488/12546}, respectively. \\
There are many different conventions for describing the diffeomorphism between $S^3$ and $SU(2)$. Given the usual definition of $S^3$ as $ \{(X,Y,Z,W) \,\,\, \mathrm{ s.t. } \,\,\,  X^2 + Y^2 + Z^2 + W^2 = 1 \}$, we consider the map:
\begin{align} 
    d: S^3 &\xrightarrow{} SU(2) \nonumber \\
    p = (X, Y, Z, W) &\mapsto g =
     \left( \begin{array}{cc}
    z_1 & z_2^* \\
    z_2 & - z_1^* ,
    \end{array} \right),
    \label{eq:Map_from_S3_to_SU(2)}
\end{align}
where $z_1 \defeq Z + i W \, , \,\, z_2 \defeq X + i Y
$ and $(\,)^*$ denotes complex conjugation. We choose the normalisation for the Maurer-Cartan forms as follows:
\begin{align}
    g^{-1}dg = -i \sigma_a \omega^a 
\label{eq:MC_form_our_conventions}
    \\
    dgg^{-1} = i \sigma_a \bar{\omega}^a,
\end{align}
where $\sigma_a$ are the Pauli matrices.
We find that the left-invariant one-forms are given by
\begin{align}
    \omega^a = \eta^{a}_{CB} X_B \mathrm{d}X_C ,
\end{align}
and the right-invariant ones read
\begin{align}
    \bar{\omega}^a = \bar{\eta}^{a}_{CB} X_B \mathrm{d}X_C ,
\end{align}
where again $\eta^{a}_{BC}$ and $\bar{\eta}^{a}_{BC}$ are the self-dual and anti-self-dual 't Hooft symbols, respectively. The vector fields dual to $\omega^a$, read:
\begin{align}
    V_a = \eta^{a}_{CB} X_B \frac{\partial}{\partial X_C }.
\end{align}
They generate right-translations and obey $[V_a,V_b] = 2 \epsilon_{abc}V_c$. Similarly, the vector fields dual to $\bar{\omega}^a$, read:
\begin{align}
    \bar{V}_a = \bar{\eta}^{a}_{CB} X_B \frac{\partial}{\partial X_C },
\end{align}
generate left-translations and obey $[\bar{V}_a,\bar{V}_b] = 2 \epsilon_{abc}\bar{V}_c$. \\
This is a global description, but we can also give a local one in terms of Euler angles by parametrising the group elements as 
\begin{align}
    g=\left(\begin{array}{cc} \vspace{0.1cm}
e^{i \frac{\alpha+\gamma}{2}} \cos \left(\frac{\beta}{2}\right) & -e^{i \frac{\alpha-\gamma}{2}} \sin \left(\frac{\beta}{2}\right) \\
-e^{-i \frac{\alpha-\gamma}{2}} \sin \left(\frac{\beta}{2}\right) & -e^{-i \frac{\alpha+\gamma}{2}} \cos \left(\frac{\beta}{2}\right)
\end{array}\right),
\end{align}
i.e. 
\begin{align}
\begin{array}{ll}\vspace{0.3cm}
Z=\cos \left(\frac{\alpha+\gamma}{2}\right) \cos \left(\frac{\beta}{2}\right), & W=\sin \left(\frac{\alpha+\gamma}{2}\right) \cos \left(\frac{\beta}{2}\right), \\
X=-\cos \left(\frac{\alpha-\gamma}{2}\right) \sin \left(\frac{\beta}{2}\right), & Y=\sin \left(\frac{\alpha-\gamma}{2}\right) \sin \left(\frac{\beta}{2}\right) .
\label{eq:Angular_coords_defn}
\end{array}
\end{align}
Using the same normalisation, we obtain:
\begin{align}
    \begin{aligned}
\omega^1 & = \frac{1}{2} (\cos \gamma \sin \beta \mathrm{d} \alpha - \sin \gamma \mathrm{d} \beta), \\
\omega^2 & =\frac{1}{2} (\sin \beta \sin \gamma \mathrm{d} \alpha + \cos \gamma \mathrm{d} \beta), \\
\omega^3 & =\frac{1}{2} ( - \cos \beta \mathrm{d} \alpha - \mathrm{d} \gamma),
\end{aligned}
\end{align}
and 
\begin{align}
\begin{aligned}
& \bar{\omega}^1 = \frac{1}{2} (\sin \alpha \mathrm{d} \beta - \cos \alpha \sin \beta \mathrm{d} \gamma), \\
& \bar{\omega}^2=\frac{1}{2}(\cos \alpha \mathrm{d} \beta + \sin \alpha \sin \beta \mathrm{d} \gamma), \\
& \bar{\omega}^3 = \frac{1}{2} (\mathrm{d} \alpha + \cos \beta \mathrm{d} \gamma),
\end{aligned}
\end{align}
where the usual ranges are $ 0 \leq \beta \leq \pi  , 0 \leq \gamma \leq 2\pi, 0 \leq \alpha \leq 4 \pi$. It is interesting to note that $w^{1,2,3} = - \sigma_{x,y,z}$, where $\sigma_{x,y,z}$ are the three one-forms employed by Eguchi and Hanson in the construction of their gravitational instanton (see \cite{EGUCHI197982}). They obey:
\begin{align}
    d \omega^a = - \epsilon_{abc} \,\omega^b \wedge \omega^c \, , \label{eq:Left_MC_eqn}\\
    d \bar{\omega}^a = - \epsilon_{abc} \, \bar{\omega}^b \wedge \bar{\omega}^c. \label{Right_MC_eqn}
\end{align}
We can use the above forms to write the metric on the unit sphere as:
\begin{align}
    g_{S^3} = \omega^a \otimes \omega^a = \bar{\omega}^a \otimes \bar{\omega}^a = (\mathrm{d} \alpha \otimes \mathrm{d} \alpha+\mathrm{d} \beta \otimes \mathrm{d} \beta+\mathrm{d} \gamma \otimes \mathrm{d} \gamma+2 \cos \beta \mathrm{d} \alpha \otimes \mathrm{d} \gamma).
\end{align}

The dual vector fields to $\omega^a$ are:
\begin{align}
    \begin{aligned}
V_1 & =2 (\frac{\cos \gamma}{\sin \beta} \frac{\partial}{\partial \alpha} - \sin \gamma \frac{\partial}{\partial \beta} - \cot \beta \cos \gamma \frac{\partial}{\partial \gamma}), \\
V_2 & =2 (\frac{\sin \gamma}{\sin \beta} \frac{\partial}{\partial \alpha} + \cos \gamma \frac{\partial}{\partial \beta} - \cot \beta \sin \gamma \frac{\partial}{\partial \gamma}), \\
V_3 & = -2 \frac{\partial}{\partial \gamma},
\end{aligned}
\end{align}
while the vector fields dual to $\bar{\omega}^a$ are given by
\begin{align}
    \begin{aligned}
& \bar{V}_1=2 (\cos \alpha \cot \beta \frac{\partial}{\partial \alpha} + \sin \alpha \frac{\partial}{\partial \beta} - \frac{\cos \alpha}{\sin \beta} \frac{\partial}{\partial \gamma}), \\
& \bar{V}_2=2 (-\sin \alpha \cot \beta \frac{\partial}{\partial \alpha} + \cos \alpha \frac{\partial}{\partial \beta} + \frac{\sin \alpha}{\sin \beta} \frac{\partial}{\partial \gamma}), \\
& \bar{V}_3= 2 \frac{\partial}{\partial \alpha} .
\end{aligned}
\end{align}
Once we identify $V_a$ with $\mathchorus{K}_{\,\, \hat{i}'}$ and $\bar{V}_a$ with $\mathchorus{K}_{\,\, \alpha'}$, these expressions allow us to calculate the matrix $\mathchorus{K}_{\,\, \hat{i}'}{}^{\alpha'}$ in coordinates.

\subsection{Instanton in Stereographic Coordinates and Angular Coordinates on $S^4$}
\label{sec:Appendix_instanton}
The connection for the standard $k=1$ instanton  with unit size centered at the origin, in regular gauge, reads (see \cite{vandoren2008lectures}):
\begin{align}
   A_{\mu}^a = \Big(\frac{1}{x^2 + 1} \Big) 2  \eta^{a}_{\mu \nu} x^\nu.
    \label{eq:Standard_instanton_Vandoren}
\end{align}
This expression is implicitly assuming that the generators satisfy the conventional algebra $[T_a, T_b] = \epsilon_{abc} T_c$. Clearly, for any other choice of generators (equation \ref{eq:Structure_constants_su(2)s}, for example), the components are scaled accordingly. \\ 
In \cite{DUFF19861}, however, the $k=1$ instanton is presented in angular coordinates, i.e. equation \ref{eq:Instanton_gauge}. We schematically show the connection between the two expressions. We should think of equation \ref{eq:Standard_instanton_Vandoren} not as a field living on $\R^4$, but as the coordinate expression of a connection on $S^4$. In other words, the $\{x^{\mu} \}$ are stereographic coordinates.
We recall that, on $S^4$, the stereographic projection map from the North and South poles read
\begin{align}
\psi_N: U_N &\rightarrow \R^4  \nonumber \\
    (a_1,...,a_{5}) &\mapsto \psi_N  (a_1,...,a_{5})= \frac{1}{1 - a_{5} } \big(a_1,...,a_{4}\big)\eqdef \big (x^N_1 , ..., x^N_4 \big)  \\
    \psi_S: U_S &\rightarrow \R^4  \nonumber \\
    (a_1,...,a_{5}) &\mapsto \psi_N  (a_1,...,a_{5})= \frac{1}{1 + a_{5} } \big(a_1,...,a_{4}\big)\eqdef \big (x^S_1 , ..., x^S_4 \big).
    \label{eq:Stereo_proj_S}
\end{align}
Note that, for simplicity and without loss of generality, we are dealing with a unit $S^4$, i.e.  $(a_1)^2 + (a_2)^2 + (a_3)^2 + (a_4)^2 + (a_5)^2 = 1$. \\
The coordinates used by \cite{DUFF19861}, on the other hand, can be read off from the metric \ref{eq:Metric_on_S4_right_inv}. They are a mixture of standard spherical coordinates and Euler angles, and, in one patch, they read\:\footnote{Note that, clearly, the metric is insensitive to exchanging labels on the coordinates. In particular, any choice of $a_{1,...,4}$ gives an explicit isomorphism between $S^3$ and $SU(2)$ as presented in section \ref{Sec:left_and_right_vect_fields}.}
\begin{align}
    a_5=\cos \mu \, \, \, , \quad a_4 = \sin \mu \cos u \cos \frac{\theta}{2} \,\,\, , \quad a_3 = \sin \mu \sin u \cos \frac{\theta}{2} \,\,\, , \nonumber \\
    a_2 = -\sin \mu \cos v \sin \frac{\theta}{2} \,\,\, , \quad a_1 = \sin \mu \sin v \sin \frac{\theta}{2},
    \label{eq_Mixed_coords_Duff}
\end{align}
where $u= (\phi + \psi)/2$, $v =(\phi - \psi)/2$ and  $ 0 \leq \theta \leq \pi  , 0 \leq \phi \leq 2\pi, 0 \leq \psi \leq 4 \pi$.
Thus we have that
\begin{align}
    x_i^S x_i^S + 1 = \frac{a_1^2 +a_2^2 +a_3^2 +a_4^2 }{(1+a_5)^2} +1 = \frac{2}{1+a_5}  = \frac{1}{\cos^2(\mu/2)}.
\end{align}
This shows that the scalar factor in equation \ref{eq:Instanton_gauge} matches the one in equation \ref{eq:Standard_instanton_Vandoren}. Regarding the equivalence between $\eta^{a}_{\mu \nu} x_S^\nu$ and the left-invariant form $\Sigma_i$, one can follow the same calculation presented in the previous section. We can use equations \ref{eq:Stereo_proj_S} and \ref{eq_Mixed_coords_Duff} to obtain a change of coordinates analogous to equation \ref{eq:Angular_coords_defn}, with an additional scaling depending on $\mu$. Then, the only difference is that an extra term, due to such scaling and proportional to the identity, will appear in the right-hand side of equation \ref{eq:MC_form_our_conventions}.

\subsection{Left-invariant vs Right-invariant}
\label{Sec:Left_inv_vs_right_inv}
The usual Fubini-Study metric reads
\begin{align}
    \mathrm{d} s^2=\left(1+\bar{q}_k q_k\right)^{-1} \mathrm{~d} \bar{q}_i \mathrm{~d} q_i-\left(1+\bar{q}_k q_k\right)^{-2} \bar{q}_i \mathrm{~d} q_i \mathrm{~d} \bar{q}_j q_j,
\end{align}
where $q_i$ are two quaternionic coordinates, and $\bar{(\cdot)}$ denotes conjugation.
With the parametrisation
\begin{align}
    q_1=\tan \chi \cos \left(\frac{1}{2} \mu\right) U, \quad q_2=\tan \chi \sin \left(\frac{1}{2} \mu\right) V,
\end{align}
where $U,V$ are unit quaternions so that $U \bar{U} = V \bar{V} = 1$, we can obtain a more familiar form of the metric. To do this, we first note that
\begin{align}
    2 U^{-1} d U=i \sigma_1+j \sigma_2+k \sigma_3, 2 \quad 2V^{-1} d V=i \Sigma_1+j \Sigma_2+k \Sigma_3,
    \label{eq:Left_inv_forms_for_Fubini-Study}
\end{align}
with $d \sigma_i = - \frac{1}{2} \epsilon_{ijk} \sigma_{j} \wedge \sigma_{k}$ and $d \Sigma_i = - \frac{1}{2} \epsilon_{ijk} \Sigma_{j} \wedge \Sigma_{k}$.
Then, we obtain
\begin{align}
    d s^2=d \chi^2+\frac{1}{4} \sin ^2 \chi\left[d \mu^2+\frac{1}{4} \sin ^2 \mu \omega_i^2+\frac{1}{4} \cos ^2 \chi\left(\nu_i+\cos \mu \omega_i\right)^2\right],
\end{align}
where $\nu_i = \sigma_i + \Sigma_i$ and $\omega_i = \sigma_i - \Sigma_i$.\\
Now, right-invariant one-forms can be defined analogously to \ref{eq:Left_inv_forms_for_Fubini-Study}:
\begin{align}
      2 d U U^{-1}=-i \tilde{\sigma}_1+-j \tilde{\sigma}_2+-k \tilde{\sigma}_3, \quad 2 d V  V^{-1} =-i \tilde{\Sigma}_1+-j \tilde{\Sigma}_2+-k \tilde{\Sigma}_3. 
\end{align}
Then, we have that again $d \tilde{\sigma_i} = - \frac{1}{2} \epsilon_{ijk} \tilde{\sigma}_{j} \wedge \tilde{\sigma}_{k}$ and $d \tilde{\Sigma}_i = - \frac{1}{2} \epsilon_{ijk} \tilde{\Sigma}_{j} \wedge \tilde{\Sigma}_{k}$, which holds with our conventions (see \ref{eq:Left_MC_eqn} and \ref{Right_MC_eqn}, up to normalisation).
Let us now consider the metric
\begin{align}
     \mathrm{d} s^2=\left(1+\bar{q}_k q_k\right)^{-1} \mathrm{~d}q_i \mathrm{~d}  \bar{q}_i-\left(1+\bar{q}_k q_k\right)^{-2} q_i \mathrm{~d}  \bar{q}_i \mathrm{~d} q_j  \bar{q}_j,
\end{align}
where the order of multiplication has been reversed. Then, all the steps that led to the result above still hold if we put tildes on $\sigma_i$ and $\Sigma_i$\footnote{In this section, we choose to distinguish right-invariant forms by using tildes because bars are used to denote conjugation. In the main text, right-invariant forms will be denoted by bars, since there is no risk of confusion there.}.
Hence, we obtain
\begin{align}
     d s^2=d \chi^2+\frac{1}{4} \sin ^2 \chi\left[d \mu^2+\frac{1}{4} \sin ^2 \mu \tilde{\omega}_i^2+\frac{1}{4} \cos ^2 \chi\left(\tilde{\nu}_i+\cos \mu \tilde{\omega}_i\right)^2\right],
\end{align}
where $\tilde{\nu}_i$ and $\tilde{\omega}_i$ are defined analogously to before.

\bibliographystyle{ieeetr}
\bibliography{Bibliography}{}

\begin{thebibliography}{10}

\bibitem{10.1063/1.525753}
R.~Percacci and S.~Randjbar‐Daemi, ``{Kaluza–Klein theories on bundles with
  homogeneous fibers. I},'' {\em Journal of Mathematical Physics}, vol.~24,
  pp.~807--814, 04 1983.

\bibitem{DUFF19861}
M.~Duff, B.~Nilsson, and C.~Pope, ``Kaluza-klein supergravity,'' {\em Physics
  Reports}, vol.~130, no.~1, pp.~1--142, 1986.

\bibitem{10.2307/1969983}
J.~Milnor, ``On manifolds homeomorphic to the 7-sphere,'' {\em Annals of
  Mathematics}, vol.~64, no.~2, pp.~399--405, 1956.

\bibitem{FREUND1985263}
P.~G. Freund, ``Higher-dimensional unification,'' {\em Physica D: Nonlinear
  Phenomena}, vol.~15, no.~1, pp.~263--269, 1985.

\bibitem{YAMAGISHI198447}
K.~Yamagishi, ``Supergravity on seven-dimensional homotopy spheres,'' {\em
  Physics Letters B}, vol.~134, no.~1, pp.~47--50, 1984.

\bibitem{Witten:1985xe}
E.~Witten, ``{GLOBAL GRAVITATIONAL ANOMALIES},'' {\em Commun. Math. Phys.},
  vol.~100, p.~197, 1985.

\bibitem{10.1063/1.529078}
R.~A. Baadhio and P.~Lee, ``{On the global gravitational instanton and soliton
  that are homotopy spheres},'' {\em Journal of Mathematical Physics}, vol.~32,
  pp.~2869--2874, 10 1991.

\bibitem{Brans:1992mj}
C.~H. Brans and D.~Randall, ``{Exotic differentiable structures and general
  relativity},'' {\em Gen. Rel. Grav.}, vol.~25, p.~205, 1993.

\bibitem{Asselmeyer-Maluga:2017tbn}
T.~Asselmeyer-Maluga and J.~Kr\'ol, ``{How to obtain a cosmological constant
  from small exotic $R^4$},'' {\em Phys. Dark Univ.}, vol.~19, pp.~66--77,
  2018.

\bibitem{Schleich_1999}
K.~Schleich and D.~Witt, ``Exotic spaces in quantum gravity: I. euclidean
  quantum gravity in seven dimensions,'' {\em Classical and Quantum Gravity},
  vol.~16, pp.~2447--2469, jan 1999.

\bibitem{CASTELLANI1984429}
L.~Castellani, L.~Romans, and N.~Warner, ``A classification of compactifying
  solutions for d =11 supergravity,'' {\em Nuclear Physics B}, vol.~241, no.~2,
  pp.~429--462, 1984.

\bibitem{Aharony_2000}
O.~Aharony, S.~S. Gubser, J.~Maldacena, H.~Ooguri, and Y.~Oz, ``Large n field
  theories, string theory and gravity,'' {\em Physics Reports}, vol.~323,
  pp.~183--386, jan 2000.

\bibitem{acharya1999branes}
B.~Acharya, J.~Figueroa-O'Farrill, C.~Hull, and B.~Spence, ``Branes at conical
  singularities and holography,'' 1999.

\bibitem{Fr__1999}
P.~Fr{\'{e}}, L.~Gualtieri, and P.~Termonia, ``The structure of multiplets in
  {AdS}4 and the complete osp(3$\vert$4){\texttimes}{SU}(3) spectrum of
  m-theory on {AdS}4{\texttimes}n0,1,0,'' {\em Physics Letters B}, vol.~471,
  pp.~27--38, dec 1999.

\bibitem{Awada:1982pk}
M.~A. Awada, M.~J. Duff, and C.~N. Pope, ``{N=8 Supergravity Breaks Down to
  N=1},'' {\em Phys. Rev. Lett.}, vol.~50, p.~294, 1983.

\bibitem{POPE1985352}
C.~Pope and N.~Warner, ``An su(4) invariant compactification of d = 11
  supergravity on a stretched seven-sphere,'' {\em Physics Letters B},
  vol.~150, no.~5, pp.~352--356, 1985.

\bibitem{Klebanov_2009}
I.~R. Klebanov, T.~Klose, and A.~Murugan, ``Ads4/cft3 -- squashed, stretched
  and warped,'' {\em Journal of High Energy Physics}, vol.~2009, pp.~140--140,
  mar 2009.

\bibitem{Coquereaux:1983kj}
R.~Coquereaux, ``{Comments about Riemannian geometry, Einstein spaces,
  Kaluza-Klein, 11-dimensional supergravity, and all that},'' 6 1983.

\bibitem{book}
T.~Asselmeyer-Maluga and C.~Brans, {\em Exotic Smoothness and Physics}.
\newblock 01 2007.

\bibitem{boyer2004einstein}
C.~P. Boyer, K.~Galicki, and J.~Kollár, ``Einstein metrics on spheres,'' 2004.

\bibitem{boyer2003einstein}
C.~P. Boyer, K.~Galicki, J.~Kollár, and E.~Thomas, ``Einstein metrics on
  exotic spheres in dimensions 7, 11, and 15,'' 2003.

\bibitem{10.2307/1971078}
D.~Gromoll and W.~Meyer, ``An exotic sphere with nonnegative sectional
  curvature,'' {\em Annals of Mathematics}, vol.~100, no.~2, pp.~401--406,
  1974.

\bibitem{nuimeprn10073}
M.~Joachim and D.~Wraith, ``Exotic spheres and curvature,'' {\em Bulletin of
  the American Mathematical Society}, vol.~45, no.~4, pp.~595--616, 2008.

\bibitem{10.2307/1999745}
A.~Derdzinski and A.~Rigas, ``Unflat connections in 3-sphere bundles over s4,''
  {\em Transactions of the American Mathematical Society}, vol.~265, no.~2,
  pp.~485--493, 1981.

\bibitem{Durán2001}
C.~E. Durán, ``Pointed wiedersehen metrics on exotic spheres and
  diffeomorphisms of s6,'' {\em Geometriae Dedicata}, vol.~88, no.~1-3,
  pp.~199--210, 2001.

\bibitem{McEnroe2016MILNORSCO}
R.~M. McEnroe, ``Milnor’s construction of exotic 7-spheres,'' 2016.

\bibitem{Bognat2018MILNORSES}
A.~Bognat, ``Milnor’s exotic spheres,'' 2018.

\bibitem{Exot_world}
J.~Sampietro and C.~Segovia, ``The exotic world of milnor’s spheres,'' 2023.

\bibitem{Nakahara:2003nw}
M.~Nakahara, {\em {Geometry, topology and physics}}.
\newblock 2003.

\bibitem{zbMATH03194988}
S.~Kobayashi and K.~Nomizu, {\em Foundations of differential geometry. {I}},
  vol.~15 of {\em Intersci. Tracts Pure Appl. Math.}
\newblock Interscience Publishers, New York, NY, 1963.

\bibitem{10.1063/1.522434}
Y.~M. Cho, ``{Higher‐dimensional unifications of gravitation and gauge
  theories},'' {\em Journal of Mathematical Physics}, vol.~16, pp.~2029--2035,
  09 2008.

\bibitem{PhysRevD.13.235}
L.~N. Chang, K.~I. Macrae, and F.~Mansouri, ``Geometrical approach to local
  gauge and supergauge invariance: Local gauge theories and supersymmetric
  strings,'' {\em Phys. Rev. D}, vol.~13, pp.~235--249, Jan 1976.

\bibitem{Rigas1978}
A.~Rigas, ``Some bundles of non-negative curvature.,'' {\em Mathematische
  Annalen}, vol.~232, pp.~187--194, 1978.

\bibitem{BOUWKNEGT201546}
P.~Bouwknegt, J.~Evslin, and V.~Mathai, ``Spherical t-duality ii: An infinity
  of spherical t-duals for non-principal su(2)-bundles,'' {\em Journal of
  Geometry and Physics}, vol.~92, pp.~46--54, 2015.

\bibitem{PhysRevD.15.1642}
R.~Jackiw, C.~Nohl, and C.~Rebbi, ``Conformal properties of pseudoparticle
  configurations,'' {\em Phys. Rev. D}, vol.~15, pp.~1642--1646, Mar 1977.

\bibitem{RevModPhys.51.461}
A.~Actor, ``Classical solutions of $\mathrm{SU}(2)$ yang---mills theories,''
  {\em Rev. Mod. Phys.}, vol.~51, pp.~461--525, Jul 1979.

\bibitem{DUFF198490}
M.~Duff, B.~Nilsson, C.~Pope, and N.~Warner, ``On the consistency of the
  kaluza-klein ansatz,'' {\em Physics Letters B}, vol.~149, no.~1, pp.~90--94,
  1984.

\bibitem{Oh:2011nv}
J.~J. Oh, C.~Park, and H.~S. Yang, ``{Yang-Mills Instantons from Gravitational
  Instantons},'' {\em JHEP}, vol.~04, p.~087, 2011.

\bibitem{10.2307/40067878}
S.~Goette, N.~Kitchloo, and K.~Shankar, ``Diffeomorphism type of the berger
  space so(5)/so(3),'' {\em American Journal of Mathematics}, vol.~126, no.~2,
  pp.~395--416, 2004.

\bibitem{10.1063/1.2358391}
V.~Gerdt, R.~Horan, A.~Khvedelidze, M.~Lavelle, D.~McMullan, and Y.~Palii,
  ``{On the Hamiltonian reduction of geodesic motion on SU(3) to
  SU(3)/SU(2)},'' {\em Journal of Mathematical Physics}, vol.~47, p.~112902, 11
  2006.

\bibitem{https://doi.org/10.15488/12546}
K.~Kumar, ``Solutions of yang–mills theory in four-dimensional de sitter
  space,'' 2022.

\bibitem{EGUCHI197982}
T.~Eguchi and A.~J. Hanson, ``Self-dual solutions to euclidean gravity,'' {\em
  Annals of Physics}, vol.~120, no.~1, pp.~82--106, 1979.

\bibitem{vandoren2008lectures}
S.~Vandoren and P.~van Nieuwenhuizen, ``Lectures on instantons,'' 2008.

\end{thebibliography}

\end{document}